\documentclass[twocolumn]{aastex6}
\submitted{Accepted for publication in AJ on October 25, 2017 }
\usepackage{graphicx}
\usepackage{hyperref}
\usepackage{natbib}
\usepackage{latexsym}
\usepackage{amssymb}
\usepackage{amsmath}
\usepackage{url}
\newcommand\kms{\hbox{km\,s$^{-1}\ $}}
\newcommand\ms{\hbox{m\,s$^{-1}\ $}}
\definecolor{aqua}{rgb}{0.0,0.67,1.0}
\hypersetup{citecolor=aqua, linkcolor=aqua}

\begin{document}

\title{Three Super-Earths Transiting the nearby star GJ 9827}
\author{Prajwal~Niraula\altaffilmark{1}, 
Seth~Redfield\altaffilmark{1}, 
Fei~Dai\altaffilmark{2,3},
Oscar~Barrag\'an\altaffilmark{4},
Davide~Gandolfi\altaffilmark{4}, 
P. Wilson Cauley\altaffilmark{1},
Teruyuki~Hirano\altaffilmark{5}, 
Judith~Korth\altaffilmark{6}, 
Alexis~M.\,S.~Smith\altaffilmark{7},
Jorge~Prieto-Arranz\altaffilmark{8,9},
Sascha~Grziwa\altaffilmark{6},
Malcolm~Fridlund\altaffilmark{10,11}, 
Carina~M.~Persson\altaffilmark{11},
Anders~Bo~Justesen\altaffilmark{12}, 
Joshua~N.~Winn\altaffilmark{2},
Simon~Albrecht\altaffilmark{12},
William~D.~Cochran\altaffilmark{13}, 
Szilard~Csizmadia\altaffilmark{7},
Girish~M.~Duvvuri\altaffilmark{1},
Michael~Endl\altaffilmark{13},
Artie~P.~Hatzes\altaffilmark{14},
John~H.~Livingston\altaffilmark{15},
Norio~Narita\altaffilmark{15,16,17},
David~Nespral\altaffilmark{8,9},
Grzegorz~Nowak\altaffilmark{8,9},
Martin~P\"atzold\altaffilmark{6},
Enric~Palle\altaffilmark{8,9}, and
Vincent~Van~Eylen\altaffilmark{10}
}

\altaffiltext{1}{Astronomy Department and Van Vleck Observatory, Wesleyan University, Middletown, CT 06459, USA; pniraula@wesleyan.edu}
\altaffiltext{2}{Department of Astrophysical Sciences, Princeton University, 4 Ivy Lane, Princeton, NJ 08544, USA}
\altaffiltext{3}{Department of Physics and Kavli Institute for Astrophysics and Space Research, Massachusetts Institute of Technology, Cambridge, MA 02139, USA}
\altaffiltext{4}{Dipartimento di Fisica, Universit\'a di Torino, via P. Giuria 1, 10125 Torino, Italy}
\altaffiltext{5}{Department of Earth and Planetary Sciences, Tokyo Institute of Technology, 2-12-1 Ookayama, Meguro-ku, Tokyo 152-8551, Japan}
\altaffiltext{6}{Rheinisches Institut f\"ur Umweltforschung an der Universit\"at zu K\"oln, Aachener Strasse 209, 50931 K\"oln, Germany}
\altaffiltext{7}{Institute of Planetary Research, German Aerospace Center, Rutherfordstrasse 2, 12489 Berlin, Germany}
\altaffiltext{8}{Instituto de Astrof\'\i sica de Canarias, C/\,V\'\i a L\'actea s/n, 38205 La Laguna, Spain}
\altaffiltext{9}{Departamento de Astrof\'isica, Universidad de La Laguna, 38206 La Laguna, Spain}
\altaffiltext{10}{Leiden Observatory, University of Leiden, PO Box 9513, 2300 RA, Leiden, The Netherlands}
\altaffiltext{11}{ Department of Space, Earth and Environment, Chalmers University of Technology, Onsala Space Observatory, 439 92 Onsala, Sweden}
\altaffiltext{12}{Stellar Astrophysics Centre (SAC), Department of Physics
and Astronomy, Aarhus University, Ny Munkegade 120, DK-8000}
\altaffiltext{13}{Department of Astronomy and McDonald Observatory, University of Texas at Austin, 2515 Speedway,~Stop~C1400,~Austin,~TX~78712,~USA}
\altaffiltext{14}{Th\"uringer Landessternwarte Tautenburg, Sternwarte 5, D-07778 Tautenberg, Germany}
\altaffiltext{15}{Department of Astronomy, University of Tokyo, 7-3-1 Hongo, Bunkyo-ku, Tokyo 113-0033, Japan}
\altaffiltext{16}{Astrobiology Center, NINS, 2-21-1 Osawa, Mitaka, Tokyo 181-8588, Japan}
\altaffiltext{17}{National Astronomical Observatory of Japan, NINS, 2-21-1 Osawa, Mitaka, Tokyo 181-8588, Japan}
\altaffiltext{}{Submitted\today, Accept \today, printed \today}

\begin{abstract} 
We report on the discovery of three transiting planets around GJ~9827. The planets have radii of 1.75$\pm$0.18, 1.36$\pm$0.14, and 2.11$ _{- 0.21 }^{+ 0.22 }$~R$_{\earth}$, and periods of 1.20896 , 3.6480, and 6.2014 days, respectively. The detection was made in Campaign 12 observations as part of our {\it K2} survey of nearby stars. GJ~9827 is a $V = 10.39$~mag K6V star at distance of 30.3 $\pm$ 1.6 parsecs and the nearest star to be found hosting planets by {\it Kepler} and {\it K2}. The radial velocity follow-up, high resolution imaging, and detection of multiple transiting objects near commensurability drastically reduce the false positive probability. The orbital periods of GJ~9827~b, c and d planets are very close to the 1:3:5 mean motion resonance. Our preliminary analysis shows that GJ~9827 planets are excellent candidates for atmospheric observations. Besides, the planetary radii span  both sides of the rocky and gaseous divide, hence the system will be an asset in expanding our understanding of the threshold.\end{abstract}


\section{INTRODUCTION} 
\label{sec:intro}
\nopagebreak
Temporal monitoring of neighboring stars (e.g., within 100 parsecs and therefore relatively bright) provides an opportunity to search for nearby planetary systems that are optimal for follow-up studies. This includes favorable conditions to characterize the system as a whole, particularly properties that can be directly linked to the planetary atmosphere and habitability, such as the stellar UV emission \citep{linsky2014}, stellar wind strength \citep{wood05} and stellar magnetic field structure \citep{gomez2016}. As the {\it Kepler} mission and ground-based radial velocity (RV) searches have shown, terrestrial planets are ubiquitous \citep{howard2012,fressin2013}.  The sample of terrestrial exoplanets will continue to grow with dedicated ground and space-based surveys \citep[e.g., {\it K2}, and in the future with the {\it Transiting Exoplanet Survey Satellite (TESS)};][]{ricker2015}.  A major scientific endeavor related to this population of planets will be the evaluation of habitability and a search for biosignatures. It is precisely in these bright, nearby systems where the atmospheric measurements will be the most sensitive, and the question of habitability will be examined in the greatest detail in the decades to come.  

\textit{K2}, the repurposed \textit{Kepler} mission,  has continued the legacy of planet discovery by its predecessor \citep{howell2014}. While the {\it K2} fields can only be monitored for about 80 days, and thereby limiting discoveries to relatively short period transiting objects, its ability to observe different parts of the ecliptic plane and choice of more diverse targets has led to some intriguing discoveries. Many planetary candidates have been reported \citep[e.g.,][]{crossfield2016} along with the first detection of transiting bodies orbiting the white dwarf  WD~1145+017 \citep{vanderburgwd1145_2015}. {\it K2} also continues to find multiplanetary systems, which are of interest for the study of planetary architecture and formation. \citet{sinukoff2016} reported the detection of eleven  multiplanetary systems from {\it K2} Campaigns 1 and 2. However, there are few  such systems around nearby stars \citep{armstrong2015, crossfield2015, gandolfi2017}, and only a handful around brighter stars that are suitable for spectroscopic characterization. 

We have detected a new planetary system hosted by a K6V star, GJ~9827 (EPIC~246389858). At 30.3 $\pm$ 1.6 parsecs, it is the nearest planetary system detected by {\it Kepler} or {\it K2}. Our analysis of the {\it Kepler} light curve demonstrates the presence of three super-Earths of radii around GJ 9827. We will use the designation of super-Earth for planets with radii from 1.25--2 R$_{\earth}$ \citep[e.g.,][]{batalha2013}, although note that the precise limits of this range are largely arbitrary and GJ~9827~d lies {\bf just above} the upper bound of this designation. The planets orbit at a distance of 0.020 $\pm$ 0.002, 0.041 $\pm$ 0.003 and 0.059 $^{+0.004}_{-0.005}$  AU corresponding to orbital periods of 1.208957$ _{- 0.000013 }^{+ 0.000012 }$, 3.64802$\pm$0.00011, and 6.20141 $ _{- 0.00010 }^{+ 0.00012 }$ days respectively. The planetary system is tightly packed, and the periods are close to 1:3:5 commensurability.  In addition to the fact that GJ 9827 is a relatively bright star, the planets occur on both sides of the rocky and gaseous threshold of $\sim$1.5 R$_{\earth}$ \citep{weiss2014, rogers2015}. Hence the system is likely to be a great asset in understanding the nature of this threshold, and could potentially exhibit a range of densities like the Kepler-36 planets \citep{carter2012}.
 
GJ~9827 planets are great candidates for atmospheric studies. In the past, ground based telescopes, along with the {\it Hubble Space Telescope} ({\it HST}) and {\it Spitzer}, have been successfully used to characterize the atmospheres of hot Jupiters \citep{charbonneau2002,knutson2008,redfield2008,sing2015}. With the {\it James Webb Space Telescope} ({\it JWST}), this territory will be extended into the super-Earth regime \citep{deming2009}. Bright, nearby planetary systems like GJ~9827, will provide excellent opportunities to probe the conditions of super-Earth atmosphere. 

\section{Observations and Data Analysis}
GJ~9827 (EPIC~246389858) was proposed by our team (PI Redfield) as part of a Campaign 12 survey of nearby stars (GO-12039), and in three other programs: GO-12071, PI Charbonneau; GO-12049 PI Quintana; and GO-12123 PI Stello. The star was observed for a total of 78.89 days from 15 December, 2016 to 4 March, 2017 at the boundary of constellation Aquarius and Pisces at RA of 23:27:04.835 and declination -01:17:10.58 in long cadence mode.

\begin{figure*}[ht!]
\centering
\includegraphics[width=1.0\textwidth, clip, trim=25mm 15mm 30mm 25mm]{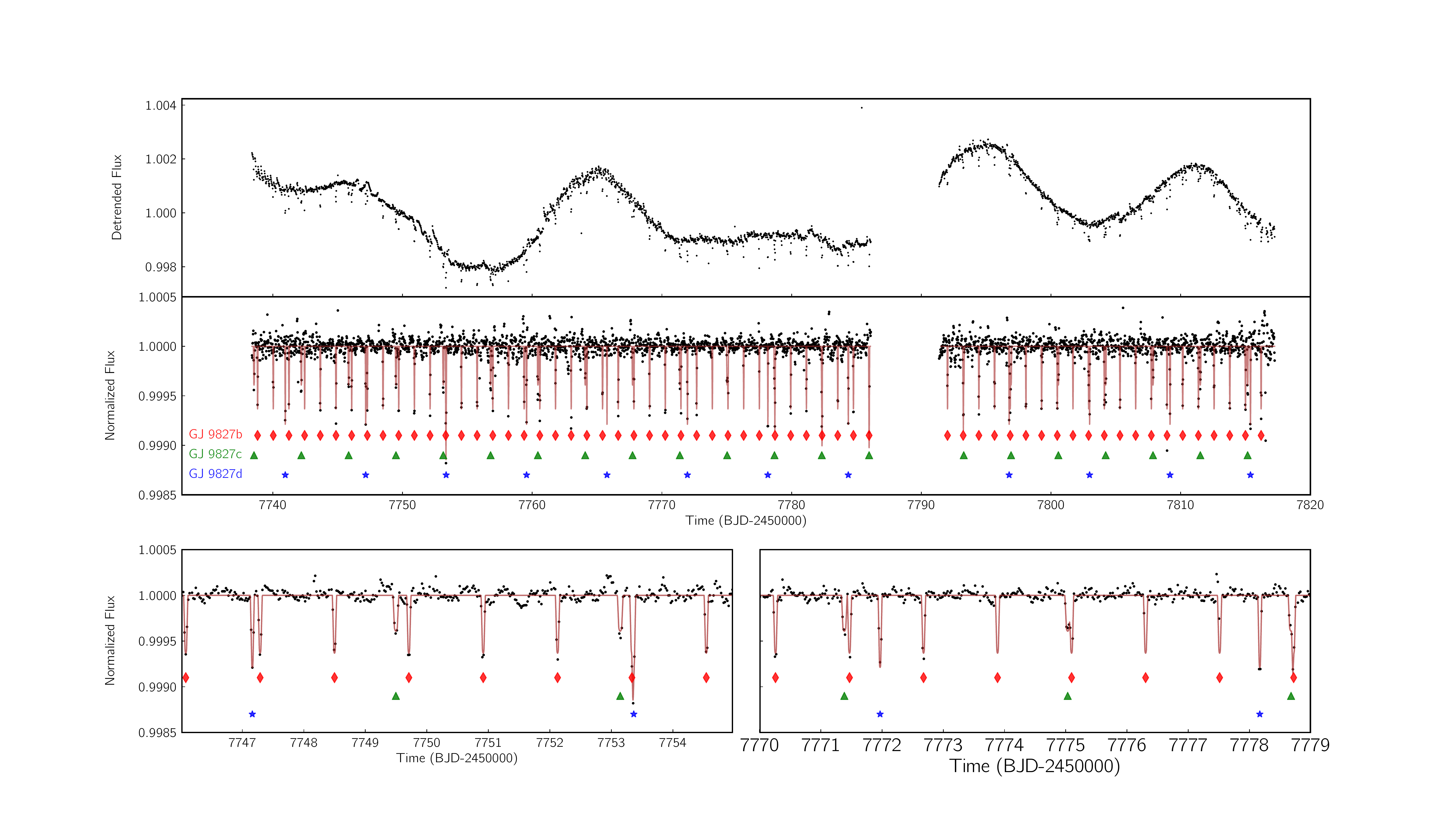} 
\figcaption{\label{fig:lightcurve} Detrended and normalized {\it K2} light curve of EPIC 246389858. Transits of each planet are marked, and the combined fit (brown line) at a finer sampling rate for all transit  based on MCMC fits, presented in \autoref{table:planetaryparams}, is shown. The bottom left and bottom right figure zooms into two different sections of the data.}
\end{figure*}

\subsection{{\it K2} Observations}

We implement a data reduction pipeline to detrend the systematic {\it K2} noise. We follow the protocol to decorrelate the data against its arclength (1D) using one of the three standard stars from the Campaign \citep[e.g.,][]{vanderburg2014, vanderburg2016}. These standard pointing stars are chosen such that their centroid can be found with better precision than an average star in the field. Among these three standards, the light curve is decorrelated with the star whose centroid variation over time is best fit with a fifth-degree polynomial, in this case EPIC~246292491. Besides, we use a modified version of \citet{vaneylen2016} publicly available code\footnote{\href{https://github.com/vincentvaneylen/k2photometry}{https://github.com/vincentvaneylen/k2photometry}}, which detrends the lightcurve by a simultaneous second order fit for both the centroid coordinates and time, also allowing for a cross term between two centroids. The {\tt k2photometry} pipeline yields a flattened light curve. In our implementation, the final transit removed light curve from {\tt k2photometry} has a standard deviation of 77 ppm compared to 106 ppm from Vanderburg's method.  Thus in \autoref{fig:lightcurve}, we show the detrended flux obtained from Vanderburg's method and the normalized lightcurve from {\tt k2photometry}. These values are higher by a factor of $\sim$2 than the expected calculated rms values of 39.2 for 10.5 V magnitude star\footnote{\href{https://keplergo.arc.nasa.gov/CalibrationSN.shtml}{https://keplergo.arc.nasa.gov/CalibrationSN.shtml}} which is expected due to pointing induced errors for \textit{K2}.

As for some of the unique aspects of our pipeline, we take the median value in each frame as the background. In order to avoid the effect of the outliers, we perform an iterative spline fitting, rejecting 3$\sigma$ outliers until convergence. Finally, the background is subtracted from the photometric flux. We reject the data with bad quality flags, which resulted in  excluding around 15\% of the data flagged for thruster firing, Agrabrightening, cosmic ray detection, and pipeline outlier detection. This has led to two instances where the transits are completely missing (refer to \autoref{fig:lightcurve}). We did a follow-up test with different aperture sizes from which a circular aperture of $\sim 20"$  radius is chosen. Initially we define our aperture as the largest contiguous region above twice the median.  From this we calculate the centroid of the star. However, the calculated centroid of the star does not coincide with the FITS coordinates probably because GJ~9827 is a high proper motion star \citep{stephenson1986}. 

Clear stellar modulation, presumably associated with stellar rotation, is evident in the detrended light curve of \autoref{fig:lightcurve}. After we remove the first five days of data which shows anomalies probably related to thermal settling, the auto correlation function \citep{mcquillan2013} of the detrended lightcurve exhibits a peak at 16.9$^{+2.14}_{-1.51}$ days, which is consistent with our reported  $v \sin i$ value of 2$\pm$1\kms assuming stellar inclination of 90$^\circ$. However, we also observed almost comparable secondary peak at 29 days, which is congruous with the value of 1.3$^{+1.5}_{-1.3}$ \kms  reported in \citet{houdebine2016}. A longer baseline of observations would help to determine the true stellar rotation period.   

We perform a Box Least-Squared \citep[BLS;][]{bls2002} search  on the flattened light curve to detect presence of any planetary signals. Once a transit signal is identified, it is fitted and removed from the light curve. In this fashion, we iteratively run the BLS algorithm on the light curve for further detection of additional transit signals. In GJ~9827, this showed a presence of three transiting planets. A simultaneous fit for all of the three identified transits is then performed with the {\tt batman} model supersampled by a factor of 15, and adjusted for {\it K2}'s long cadence \citep{kriedberg2015}. We use the affine invariant MCMC method implemented in {\tt emcee} \citep{emcee} with 100 walkers for 30000 steps; of this, the first 22500 steps were removed as burn-in. The rest of the data is used to build the posterior distributions and estimate the uncertainties in our transit  parameters.

\begin{figure*}[ht!]
\centering
\includegraphics[width=\textwidth, scale=1]{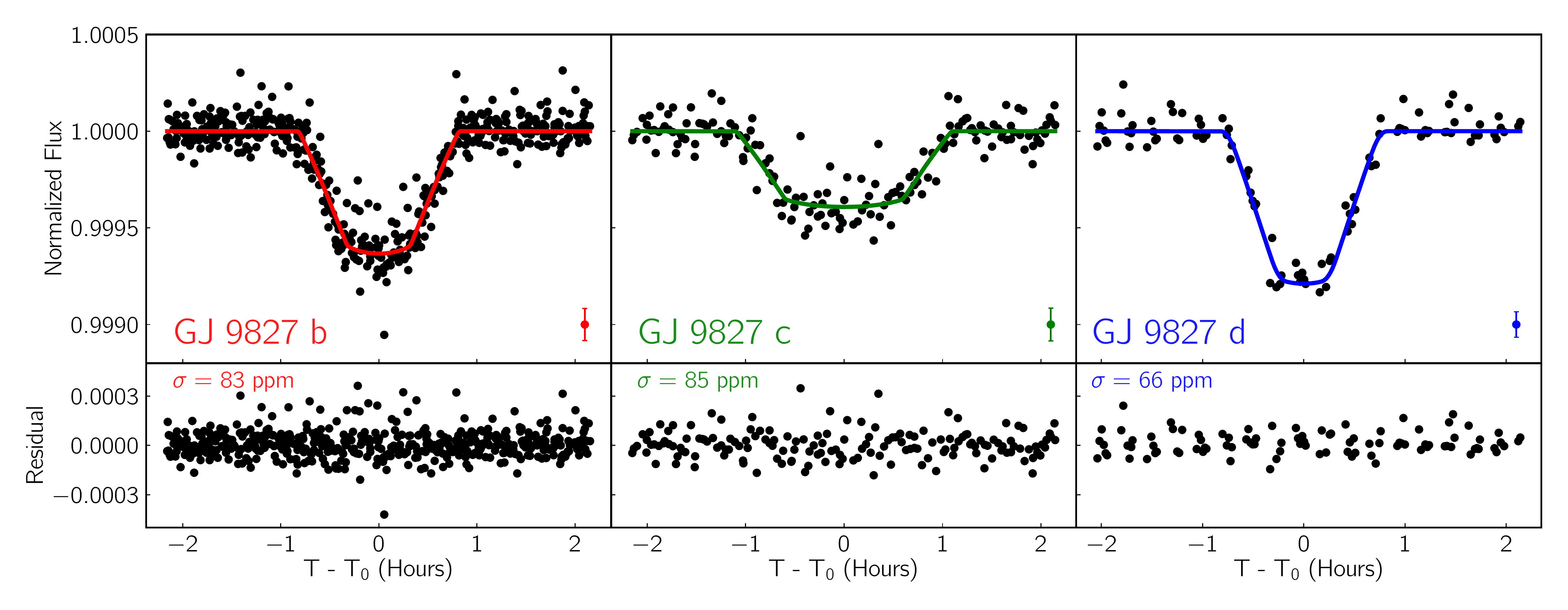} 
\figcaption{\label{fig:transit}Model Fit of MCMC obtained parameters for GJ~9827~b, GJ~9827~c, and GJ~9827~d. The parameters are available in \autoref{table:planetaryparams}.  Note the normalized flux scale is kept constant for comparison. 1$\sigma$ error bars \textbf{computed from the respective residuals} are shown in the right hand bottom corner for reference.}
\vspace*{0.01in}
\end{figure*}

We use uniform priors for the period, time of conjunction, scaled planet radius and impact parameter for all three planets. For limb darkening parameters, we use triangular sampling suggested by \citet{kipping2013}. We additionally use \citet{sing2010} to introduce Gaussian priors on limb darkening based on the stellar parameters. We use mean value of 0.5782 for u$_1$, and 0.1428 for u$_2$, both with 0.1 standard deviation. Since this is a short period multi-planetary system, we assume tidal circularization of the orbits and adopt a fixed eccentricity of $e=0$ for all three planets \citep{vaneylen2015}. As for the scaled semi-major axis of GJ~9827~c and d, we assume they are constrained by Kepler's Third Law. As a result, we fit 15 independent variables (\autoref{table:planetaryparams}). We additionally introduce a Gaussian prior based on the spectroscopically derived stellar density  of 3.37$\pm$0.51 g cm$^{-3}$. MCMC runs without Gaussian priors on sometimes converged to unrealistic semi-major axis values, hence the choice. From the posterior distribution, most of the variables are well constrained except for limb darkening parameters. Due to short transit duration and long integration time  for {\it K2} , limb darkening parameters are not expected to be well constrained \citep{kipping2010}. The introduction of Gaussian prior for limb darkening parameters does not noticeably affect the other fit parameters.

It is interesting to note that the transit duration is longest for GJ~9827~c, and shortest for GJ~9827~d. This is consistent with the fit's prediction that GJ~9827~d has a higher impact parameter than either  GJ~9827~b or c. Additional independent MCMC runs were performed by our team using {\tt pyaneti} \citep{pyaneti},  with flattened lightcurves from independent pipelines developed in our group, and the results are within 1$\sigma $ errors. Note that the high impact parameter of GJ~9827~d suggests  additional planets, if present, are likely to be non-transiting. This possibility will be explored in the follow-up RV campaign.

\begin{deluxetable*}{lcccc}
\setlength{\tabcolsep}{12pt}
\setlength{\tabcolsep}{12pt}
\tablecaption{\label{table:planetaryparams} Planetary  parameters of GJ~9827~b, c and d.}
\tablehead{
\colhead{Parameter } &
\colhead{Unit} &
\colhead{GJ~9827~b} & 
\colhead{GJ~9827~c} & 
\colhead{GJ~9827~d} 
}
\startdata 
\rule{0pt}{4ex}Transit Epoch BJD--2450000 ($T_0$) & day&7738.82671 $_{- 0.00046 }^{+ 0.00043} $&  7738.5519 $_{- 0.0014}^{+ 0.0014} $&7740.96100 $_{- 0.00087}^{+ 0.00083}$\\
Period ($P_{\rm{orb}})$ & day & 1.208957 $_{- 0.000013 }^{+ 0.000012 }$&3.64802 $_{- 0.00011 }^{+ 0.00011 }$&6.20141 $_{- 0.00010 }^{+ 0.00012}$ \\
Scaled planet radius ($R_p/R_*$)&- & 0.0246 $ _{- 0.0005}^{+ 0.0003}$& 0.0192 $_{- 0.0005 }^{+ 0.0004}$ & 0.0297 $_{- 0.0008 }^{+ 0.0010}$\\
Scaled Semimajor axis ($a/R_*$)&-& 6.55 $_{- 0.32}^{+ 0.30}$& 13.67 $ _{-0.63}^{+ 0.66}$ & 19.5 $_{- 0.90}^{+0.95}$\\
Impact Parameter ($b$)&-& 0.595 $_{- 0.070}^{+ 0.056}$& 0.558 $_{- 0.096}^{+ 0.068}$ & 0.910 $_{- 0.013}^{+ 0.011}$\\
\textbf{Derived Parameters}&\\
Planet Radius ($R_p$) & R$_\earth$ & 1.75 $_{- 0.18}^{+ 0.18}$& 1.36 $_{- 0.14}^{+ 0.14}$&2.11 $_{- 0.21}^{+ 0.22}$\\
Semi Major Axis ($a$) & AU & 0.020 $ _{- 0.002}^{+ 0.002}$ & 0.041 $_{-0.003}^{+0.003}$ & 0.059 $_{-0.005}^{+0.004}$\\
Transit Duration ($T_{14}$) & hour & 1.12 $_{-0.07}^{+0.06}$ & 1.69 $_{+0.10}^{-0.11}$ & 1.01 $_{-0.05}^{+0.05}$\\
Orbital Inclination ($i$)&$ \deg$ & 84.86$_{-0.54}^{+0.54}$& 87.66 $_{-0.31}^{+0.30}$ & 87.32 $_{-0.13}^{+0.12}$\\
Equilibrium Temperature ($T_{\rm{eq}}$)&K&1075$_{-37}^{+38}$$^a$ & 744 $_{-26}^{+26}$$^a$ & 623$_{-22}^{-22}$$^a$\vspace{0.2cm}\\
\hline
\hline
\rule{0pt}{4ex}\textbf{Limb Darkening Coefficients}&&\\
u$_1$&-&0.35 $ _{- 0.07}^{+ 0.07}$ $^b$\\
u$_2$&-& 0.00$ _{- 0.13 }^{+ 0.23 }$
 $^b$\vspace{0.2cm}\\
\enddata
\tablenotetext{}{Note: The values of eccentricity for all three planets is fixed at zero.}
\tablenotetext{a}{We calculate equilibrium temperature as T$_{\rm{eq}}$ = T$_*\sqrt{R_*/2a}(1- \alpha)^{1/4}$, where Bond Albedo ($\alpha$) is adopted at 0.3.}
\tablenotetext{b}{A single set of limb darkening parameters is fitted for three different transit light curves.}
\end{deluxetable*}

\subsection{Spectroscopic Observations}

We collected seven high-resolution (R\,$\approx$\,67,000) spectra of GJ~9827 using the FIbre-fed \'Echelle Spectrograph \citep[FIES;][]{Frandsen1999,Telting2014} mounted at the 2.56 m Nordic Optical Telescope (NOT) of Roque de los Muchachos Observatory (La Palma, Spain). The follow-up was performed between 20 July and 1 August 2017 UT as part of the OPTICON observing program 2017A/064, under clear and stable weather conditions, with seeing ranging between $0{\farcs}5$ and $0{\farcs}8$. For each observation epoch, we took 3 consecutive sub-exposures of 900 seconds that were average combined using a sigma-clipping algorithm to remove cosmic ray hits. Following the observing strategy described in \citet{Buchhave2010} and \citet{Gandolfi2013}, we traced the RV drift of the instrument by acquiring ThAr spectra with long exposure ($T_\mathrm{exp}$\,=\,65\,sec) taken immediately before and after each observation. We reduced the FIES data using standard IRAF and IDL routines, which include bias subtraction, flat fielding, order tracing and extraction, and wavelength calibration. RV measurements were extracted using multi-order cross-correlation technique with the RV standard star HD~190007 -- observed with the same instrument set-up as the target object -- for which we adopted a heliocentric RV of $-30.40$\,\kms, as measured by \citet{Udry1999}. We report the RVs and their uncertainties in \autoref{table:rvdata}. Our measurements show no significant RV variation: the rms is 2.4\,\ms, which is comparable to the mean nominal uncertainty of 3.1\,\ms. 	

We used the co-added FIES spectrum, which has a SNR ratio of $\sim$150 per pixel at 5500\,\AA, to derive the fundamental parameters of GJ~9827. The analysis was performed following the procedures already adopted for other \textit{K2} host stars \citep{barragan2017, fridlund2017, gandolfi2017, guenther2017, johnson2016}. We took advantage of four different spectral analysis packages applied independently by different sub-groups within our team. The four analyses provide consistent results well within the errors bars. While we have no strong reason to prefer one method over the other, we adopted the results obtained using {\tt SpecMatch-Emp} \citep{yee2017}.  This technique relies on the use of high-resolution template spectra of stars whose effective temperature ($T_{\mathrm{eff}}$), radius ($R_\star$), and iron abundance ([Fe/H]), have been accurately measured by interferometry, spectrophotometry, and spectral synthesis. We use \citet{mann2015}'s empirical relations to derive the stellar mass. Our stellar parameters are presented in \autoref{table:stellarparams}. The values are consistent with those reported by \citet{houdebine2016}.

\begin{table}[!t]
\begin{center}
\caption{FIES RV measurements.}
\label{table:rvdata}
\label{table:rvdata}
\begin{tabular}{ccc}
\hline
\hline
\noalign{\smallskip}
BJD$_\mathrm{TDB}$ & RV  & Error  \\
 $-2,450,000$ & (\kms) & (\kms) \\ 
\noalign{\smallskip}
\hline
\noalign{\smallskip}
7954.617085 & 31.7746 & 0.0033 \\
7955.612895 & 31.7724 & 0.0032 \\
7956.627456 & 31.7751 & 0.0025 \\
7964.582846 & 31.7796 & 0.0028 \\
7965.593839 & 31.7739 & 0.0032 \\
7966.573354 & 31.7728 & 0.0033 \\
7966.707233 & 31.7735 & 0.0035 \\
\noalign{\smallskip}
\hline
\end{tabular}
\end{center}
\end{table}

\subsection{Limits on a stellar companion}

We investigate the probability that the transit signals are of a non-planetary origin coming from a background source or a companion. \citet{lissauer2012} estimate the false positive probability for systems with three transiting planet candidates at $<0.4\%$ with the extremely conservative assumption of a 50\% false positive rate for single planet candidates. The probabilities for detecting 1 planet$+$2 false positives or 2 planets$+$1 false positive are even lower than the 3 planet case. The short orbital periods also argue against a massive triple system, which would be dynamically unstable and produce very large transit timing variations (TTV) \citep{lissauer2011}, which are not observed (see Section 3.1). 

Using high-resolution Lucky Imaging \textit{I}-band observations, \citet{jodar2013} find no evidence of a stellar companion to GJ~9827. They rule out all companions with $T_{\mathrm{eff}} \gtrsim 3200$ K, or earlier than spectral type M4, at angular separations $\gtrsim 0.5''$. The constraints are even tighter for angular separations $\gtrsim 1.0''$, ruling out all companions with $T_{\mathrm{eff}} \gtrsim 2800$ K, or earlier than spectral type $\sim$M6.5.  

We can also rule out companions with $T_{\mathrm{eff}} \gtrsim 3200$ at any separation by assuming normal main sequence dwarf parameters \citep{pecaut13}: if a bound, unresolved companion is present, anything with $T_{\mathrm{eff}} \gtrsim 3000$ K would result in $J \lesssim 7.95$. This is incompatible with the measured 2MASS J-magnitude of $J = 7.984 \pm 0.020$ \citep{skrutskie06}. Thus, any undetected bound stellar companions to GJ 9827 must have a spectral type later than $\sim$M5. Using optical and infrared photometry, including the Wide-Field Infrared Survey Explorer (WISE) 3.4, 4.6, 12.0, and 22.0 $\mu$m magnitudes \citep{wright2010}, we also find no evidence of any infrared excess. Although we cannot rule out the existence of a faint late-type companion, we currently favor GJ~9827 as the host star. We note that the planetary radii necessary to produce the observed transits depths are still $\sim$3-5 $R_\earth$ if the candidates orbit an undetected late-type companion, placing them in the mini-Neptune regime. Follow-up RV observations and high-contrast adaptive optics imaging will help confirm the nature of the planets' parent star.
\begin{figure*}[ht!]
\centering
\includegraphics[width=1.0\textwidth ]{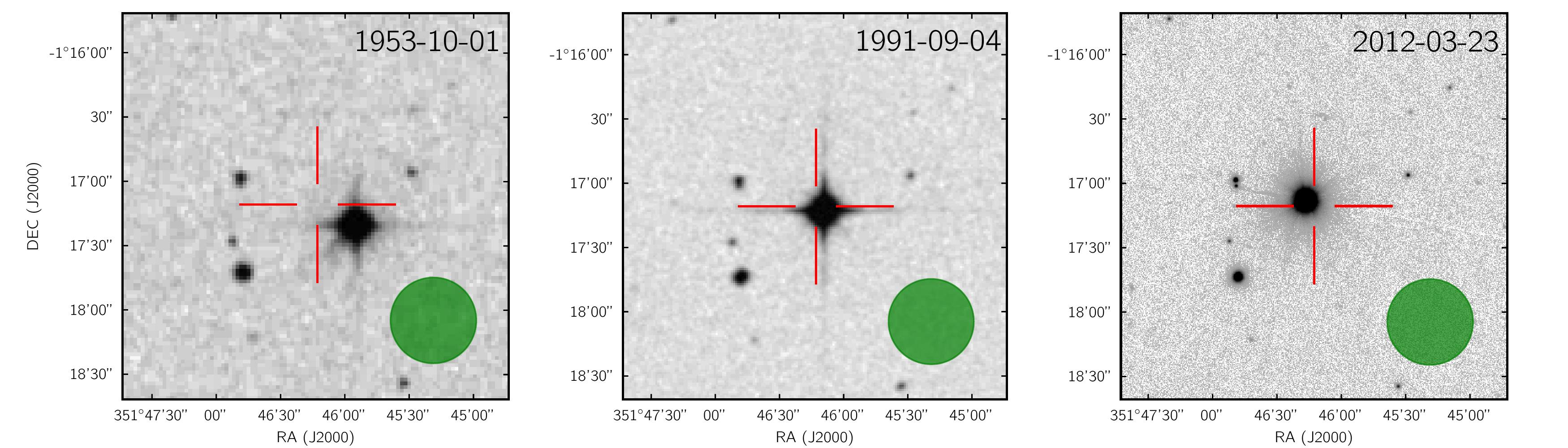} 
\figcaption{\label{fig:arxivalimage} Archival image in r band of GJ~9827 from the POSS-I and II from year 1953 and 1991. The third image is from more recent  Pan-STARRS in g band from year 2012. No background objects concurrent with current position of GJ~9827 is seen in the archival image. The green circle in each image shows 20" aperture size used for \textit{K2} photometry, meanwhile the reference position of GJ~9827 at the J2000 epoch is indicated with a red reticle.}
\end{figure*}

Given its large proper motion ($\approx 400 ~\mathrm{mas~yr}^{-1}$), we  are able to rule out the possibility of an unbound background contamination using the archival data. Using the STScI Digitized Sky Survey\footnote{\href{http://stdatu.stsci.edu/cgi-bin/dss\_form}{http://stdatu.stsci.edu/cgi-bin/dss\_form}}, we identify GJ~9827 images as early as 1953 (see \autoref{fig:arxivalimage}). By comparing the image to the latest epoch (2012), we determine that there is no background object coincident with its current position visible in the 1953 plate. In order to estimate the limiting magnitude of the 1953 image, we considered an object near to our target which is faint, but clearly above the detection threshold of the image. By reference to the SDSS catalog, we determined that this object has $r = 19.0$ (cf. $R = 10.1$ for GJ~9827). We, therefore, conclude that the 1953 plate is sensitive to objects about 9 magnitudes fainter than GJ~9827, and we can rule out the presence of unbound contaminants brighter than this. An equal mass eclipsing binary system with a combined magnitude of $r=19.0$ would produce at most a 125~ppm deep signal in the light curve of GJ~9827, which is shallower than the observed transits.

\begin{deluxetable}{lcc}
\tablecaption{\label{table:stellarparams} Stellar Parameters of GJ~9827 (EPIC 246389858)}
\tablehead{
\colhead{Parameter} &
\colhead{Units} &
\colhead{Value} 
}
\startdata
$V$ mag & - &10.39$^{\rm{a}}$\\
$J$ mag & - & 7.984$^{\rm{b}}$\\
Distance & pc&  30.3 $\pm$ 1.6$^{\rm{c}}$\\
Spectral Type &-&K6V$^{\rm{d}}$\\
Effective Temperature ($T_{\rm{eff}})$ & K & 4255 $\pm$ 110$^{\rm{d}}$\\
Surface gravity ($\log g$) & cgs & 4.70 $\pm$ 0.15 $^{\rm{d}}$ \\
Iron Abundance ([Fe/H]) & dex & -0.28$\pm$0.12$^{\rm{d}}$\\
Radius ($R_*$)  &R$\sun$&0.651 $\pm$ 0.065$^{\rm{d}}$\\
Mass ($M_*$) &M$\sun$&0.659 $\pm$ 0.060$^{\rm{d}}$\\
%Ls&L$_\odot$&0.125 $\pm$ 0.027$^{\rm{d}}$\\
$v \sin i$ & \kms & 2 $\pm$ 1$^{\rm{d}}$\\
Rotational Period ($P_{\rm{rot}}$) &day&  16.9$^{+2.14}_{-1.51}$$^{\rm{d}}$ \\
\enddata
\tablenotetext{a}{Adopted from \citet{zacharias2013}}
\tablenotetext{b}{Adopted from \citet{cutri2003}}
\tablenotetext{c}{$Hipparcos$ \citep{vanleeuwen2007}}
\tablenotetext{d}{This work}
\end{deluxetable}

\section{DISCUSSION}
\subsection{A Closely Packed super-Earth System}
Multi-transiting planetary systems offer more than conventional ways for characterizing the systems. Through transit timing variations (TTV) and transit duration variation (TDV), planetary masses and orbital elements in these systems can be constrained to higher precision than  single transiting systems \citep{agol2005, ragozzine2010}. In addition, they provide opportunity to test in-situ vs. ex-situ planetary formation, which continues to be a topic of debate in the regime of super-Earths \citep{chiang2013, schlichting2014, dangelo2016}.  

\begin{figure}[hbp]
\centering
\includegraphics[width=0.5\textwidth,clip,trim=10mm 10mm 10mm 20mm]{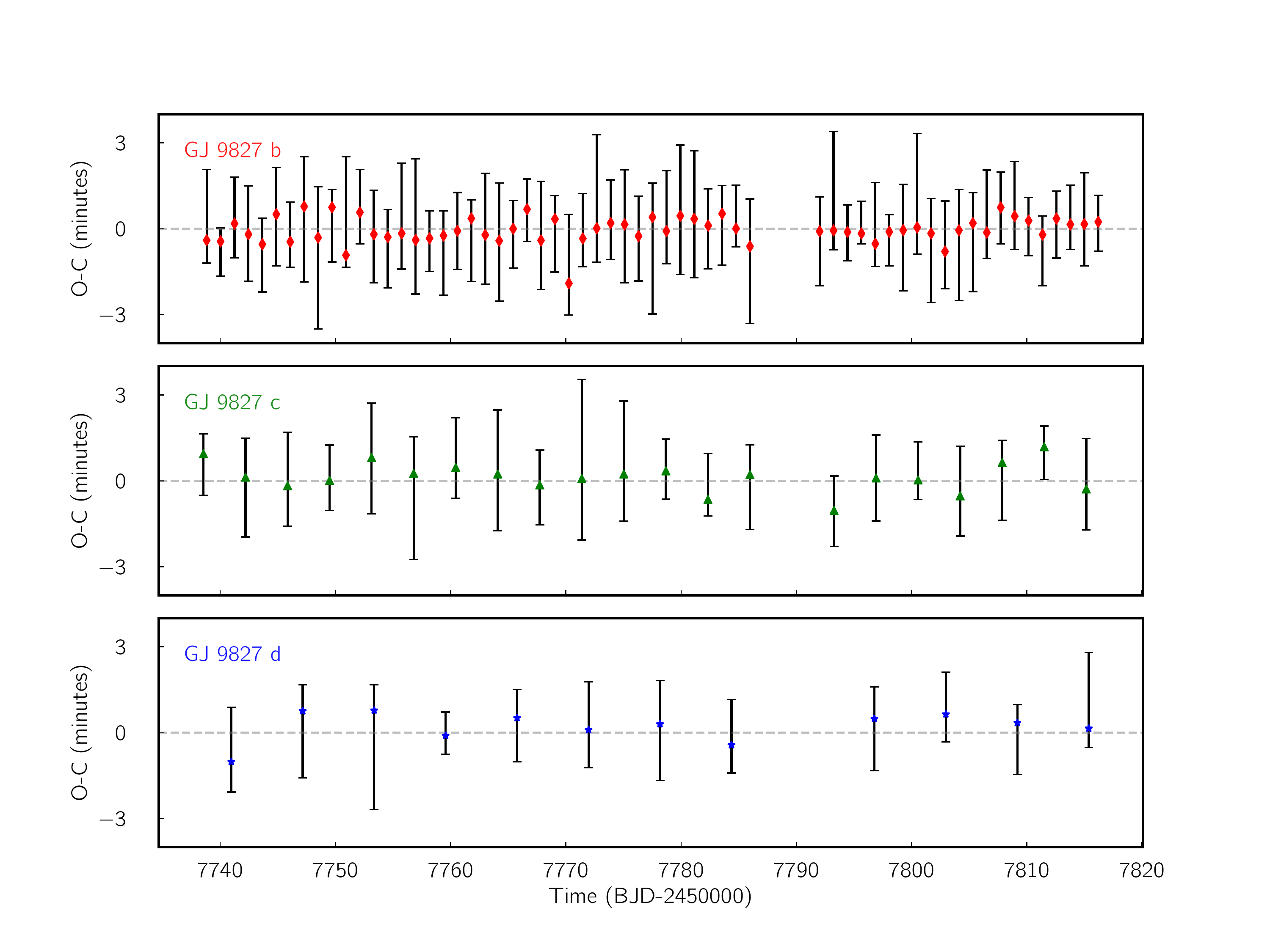} 
\figcaption{\label{fig:OCFigure}O-C Diagram for GJ~9827~b, c and d. The O-C signal and errors are estimated using MCMC fit using model created with transit parameters. No significant TTVs greater than three minutes is detected.}
\end{figure}

No TTV greater than 3 minutes were found for the planets GJ~9827~b, c, and d as shown in \autoref{fig:OCFigure}. An order of magnitude calculation of the expected TTV amplitude, based on work by \citet{agol2005}, indicates that the expected amplitude of TTVs is smaller than 3 minutes. Occurring near commensurability of 1:3:5, GJ~9827~c and b period ratio deviate from 3:1 ratio by +0.5\%, whereas period of GJ~9827~d and c deviates 5:3 by +2.0\%. Such small positive deviation from the exact resonance has been reported in other {\it Kepler} multiple planet systems \citep{fabrycky2014}. In fact, the period ratio of GJ~9827~c and d is 1.69994 $\pm$ 0.00003 ($\sim$1.7),  where \citet{steffen2015} reported the presence of a modest peak in their sample of {\it Kepler} multiple planet systems.  Examples of second order resonances in our own solar system, as well as  in exoplanetary architectures have motivated a dynamical explanation regarding their origin \citep{mustill2011, xu2017}, and a dynamical study of GJ 9827 could be useful in answering questions pertaining to such architecture.  

We also phase folded and binned the transit removed data at the period of the first planet to investigate the presence of a phase curve or of a secondary eclipse. None were evident as the overall noise in the light curve is too dominant to make any statistically significant claim. The GJ~9827 planets may be excellent candidates for searching for such signals in the infrared. 

Detected phase curves and secondary eclipse, combined with TTV observations, could help to determine the orbital and planetary parameters with greater precision. The  estimated mass of the GJ~9827 super-Earths based on the mass-radius relation proposed by \citet{weiss2014}: $M_p/M_\earth = 2.69(R_p/R_\earth)^{0.93}$ are 4.5, 3.5, and 5.4 M$_\earth$. Based on these mass estimates  and orbital parameters, the semi-amplitude of RV signals of the three planets are 3.5, 1.9 and 2.5 \ms. The threshold of 1.5 R$_\earth$, as proposed by \citet{weiss2014},  suggests GJ~9827~c to be a rocky, and GJ~9827~d to be a gaseous planet. As for GJ~9827~b, its radius lies close to the boundary itself, and in the light that the exact value of the threshold is not well known \citep{lopez2014, rogers2015, weiss2014}, we expect RV follow-up to shed more light on its density.  Details of a concentrated RV campaign will be discussed in a future paper.

\subsection{Prospects for Atmospheric Characterization}

\begin{figure}[hbp!]
\centering
\includegraphics[width=0.5\textwidth, clip, trim=0mm 0mm 20mm 10mm]{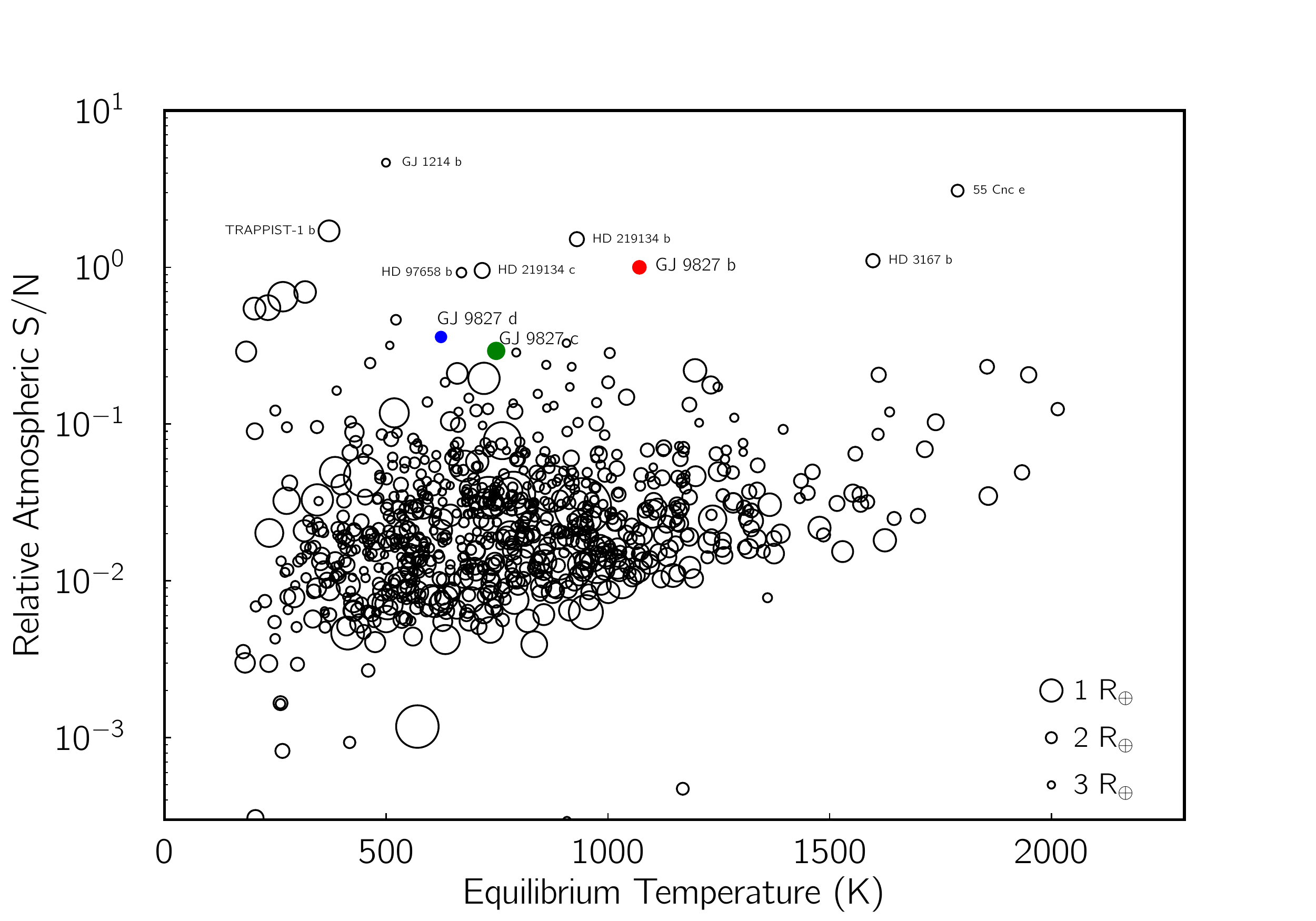} 
\figcaption{\label{fig:AtmSNR} Relative S/N ratio of an atmospheric signal for all exoplanet candidates with $R < 3 R_\oplus$. The GJ~9827 planets are the filled colored symbols with GJ~9827~b used as the S/N reference. Using this metric, GJ~9827~b is ranked as the sixth most favorable super-Earth for atmospheric characterization.}
\end{figure}
Atmospheric characterization provides an opportunity to not only measure the current conditions in the planetary atmosphere, but also put constraints on formation history and interior structure \citep{owen1999}, interactions with host star \citep{cauley2017}, atmospheric and planetary evolution \citep{oberg2011}, and biological processes \citep{meadows2010}. The planets in the GJ~9827 system offer excellent opportunities to characterize their atmospheres. \autoref{fig:AtmSNR} displays a relative atmospheric detection S/N metric (normalized to GJ~9827~b) all well characterized with $R_p < 3 R_\oplus$. The sample of small exoplanets, totaling 789\footnote{as of 15 September, 2017}, is taken from the NASA Exoplanet Archive\footnote{\href{https://exoplanetarchive.ipac.caltech.edu}{https://exoplanetarchive.ipac.caltech.edu}}. The atmospheric signal is calculated in a similar way to \citet{gillon2016} with an effective scale height \citep[$h_{\rm{eff}}$ = $7H$;][]{miller2009} using the equilibrium temperature, a Bond albedo of $\alpha = 0.3$, and an atmospheric mean molecular weight $\mu = 20$. However, since we calculate the relative signal and assume identical properties for all atmospheres, these values do not affect our results but are included for completeness. The atmospheric signal is dominated by the atmospheric scale height, favoring hot, extended atmospheres, and the host star radius, favoring small, cool stars. The relative S/N calculation scales the atmospheric signal with the properties that make it possible to detect and measure this signal,
\begin{subequations}
\begin{align}
\frac{S/N_{ }}{S/N_{\rm{Ref}}} &= \frac{W_{ }}{W_{\rm{Ref}}}\sqrt{10^{-0.4(J-J_{\rm{Ref}})}}\sqrt{\frac{P_{\rm{Ref}}T_{14}}{PT_{14_{  \rm{Ref}  }}}},\\
W &= \frac{2 R_p h_{\rm{eff}}}{R_*^2}.
\end{align}
\end{subequations}
We use the $J$-band flux \citep[e.g., H$_2$O measurements with {\it JWST};][]{beichman2014}, and scale by the duration of the transit and the frequency of transits. Given that sensitive atmospheric observations will likely require many transits to build sufficient signal \citep[e.g.,][]{cowan2015}, we have used a metric that optimizes the S/N over a period of time rather than a per-transit metric. 

Out of this sample of super-Earth exoplanets, all three planets in the GJ~9827 system are in the top 20 in terms of the S/N for atmospheric characterization. This is mainly a consequence of the brightness of this nearby cool, small, star. This highlights the powerful impact nearby stars have on exoplanet characterization given the relative brightness of even small host stars, providing strong atmospheric signals at high S/N. Using this metric, GJ~9827~b is ranked the 6th best target for atmospheric characterization, after GJ~1214~b, 55~Cnc~e, TRAPPIST-1~b, HD~219134~b, and HD~3167~b. Given that all three of the GJ~9827 planets are near commensurability, there are regular opportunities to observe two, or even all three transits at approximately the same time. For example, see the {\it K2} signal at BJD 2457753, which occurs on average every 150 days (assuming 6 hours of observation). The wait is  shorter for simultaneous transits of two planets. Transit overlap occurs for GJ~9827~b and c over 6 hours of observation on average every 8.7 days; for GJ~9827~c and d around 53 days, and for GJ~9827~b and d around 15 days.

\section{CONCLUSION}
Super-Earths are intrinsically interesting objects, as universally abundant despite being absent from our solar system. Hosting at least three super-Earths, GJ~9827 lies at a distance of a mere 30 parsecs, the closest planetary system discovered by {\it Kepler} or {\it K2}.  The planets occur on the both side of rocky gaseous divide, therefore are likely to have different range of densities and provide a test of the precise location of this division. Its three body second order resonant system is also intriguing from  the viewpoint of planetary architecture and formation. In addition, GJ~9827 is an excellent candidate for follow-up atmospheric characterization with {\it JWST} and other facilities. All these exciting features mean GJ~9827, like other nearby planetary systems around bright stars, will be a great asset for exploring the most fundamental questions of our field.

\vspace{0.5cm}

{\bf Acknowledgments:}
We are extremely grateful to the NOT staff members for their unique and superb support during the observations. 
The research leading to these results has received funding from the European Union Seventh Framework Programme (FP7/2013-2016) under grant agreement No. 312430 (OPTICON). Based on observations obtained with the Nordic Optical Telescope (NOT), operated on the island of La Palma jointly by Denmark, Finland, Iceland, Norway, and Sweden, in the Spanish Observatorio del Roque de los Muchachos (ORM) of the Instituto de Astrof\'isica de Canarias (IAC). This paper includes data taken by {\it Kepler}. Funding for the Kepler mission is provided by the NASA Science Mission directorate through grant \rm{14-K2G01\_2-0071}, submitted in response to NNH14ZDA001N Research Opportunities in Space and Earth Science (ROSES-2014). S. Redfield and P. W. Cauley acknowledge the support from the National Science Foundation through Astronomy and Astrophysics Research Grant AST-1313268. D.\,Gandolfi acknowledges the financial support of the \emph{Programma Giovani Ricercatori -- Rita Levi Montalcini -- Rientro dei Cervelli (2012)} awarded by the Italian Ministry of Education, Universities and Research (MIUR). T. Hirano acknowledges support from JSPS KAKENHI Grant Number 16K17660. S. Albrecht and A. B. Justesen acknowledge support by the Danish Council for Independent Research, through a DFF Sapere Aude Starting Grant nr. 4181-00487B. We also thank the referee for the comments and suggestions that have helped to make this paper better. 

During referee paper review process we became aware of the  similar discovery paper by \citet{rodriguez2017}.

\software{\tt batman} \citep{kriedberg2015}, {\tt emcee} \citep{emcee}, {\tt IRAF} \citep{tody1986, tody1993}, {\tt k2photometry} \citep{vaneylen2015}, {\tt matplotlib} \citep{matplotlib}, {\tt pyaneti} \citep{pyaneti}, { {\tt SpecMatch-Emp} \citep{yee2017}.}

\end{document}